\documentclass[acmsmall]{acmart}
\pdfoutput=1

\usepackage{soul}

\AtBeginDocument{%
  \providecommand\BibTeX{{%
    \normalfont B\kern-0.5em{\scshape i\kern-0.25em b}\kern-0.8em\TeX}}}

\setcopyright{rightsretained}

\begin{document}

\vspace{3pt}

\acmJournal{PACMHCI}
\acmYear{2022} \acmVolume{6} \acmNumber{CSCW1} \acmArticle{62}
\acmMonth{4} \acmPrice{} \acmDOI{10.1145/3512909}

\makeatother

\title{Project IRL: Playful Co-Located Interactions with Mobile Augmented Reality}

\title{Project IRL: Playful Co-Located Interactions with Mobile Augmented Reality}

\author{Ella Dagan}
\affiliation{%
  \institution{University of California, Santa Cruz, USA}
    \city{Sant Cruz}
  \state{CA}
  \postcode{95064}
}
\email{ella@ucsc.edu}

\author{Ana Mar{\'i}a C{\'a}rdenas Gasca}
\email{acardenasgasca@ucsb.edu}
\affiliation{%
  \institution{University of California, Santa Barbara, USA}
  \city{Santa Barbara}
  \state{CA}
  \postcode{93106}
}

\author{Ava Robinson}
\email{arobinson@snap.com}
\affiliation{%
  \institution{Snap Inc., Northwestern University, USA}
  \city{New York}
  \state{NY}
  }

\author{Anwar Noriega}
\email{anwar@wabisabi.design}
\affiliation{%
  \institution{Wabisabi Design Inc., Mexico}
  \city{Mexico City}
}

\author{Yu Jiang Tham}
\email{yujiang@snap.com}
\affiliation{%
  \institution{Snap Inc., USA}
  \city{Seattle}
  \state{WA}
  \postcode{98121}
}

\author{Rajan Vaish}
\email{rvaish@snap.com}
\affiliation{%
  \institution{Snap Inc., USA}
  \city{Santa Monica}
  \state{CA}
  \postcode{90405}
}

\author{Andr{\'e}s Monroy-Hern{\'a}ndez}
\email{amh@snap.com}
\affiliation{%
  \institution{Snap Inc., Princeton University, USA}
  \city{Seattle}
  \state{WA}
  \postcode{98121}
}

%\acmJournal{PACMHCI}
%\acmYear{2022} \acmVolume{6} \acmNumber{CSCW1} \acmArticle{62} \acmMonth{4} \acmPrice{}\acmDOI{10.1145/3512909}

%\usepackage{etoolbox}
\makeatletter
\patchcmd{\maketitle}{\@copyrightpermission}{
\begin{minipage}{0.4\columnwidth}
\href{http://creativecommons.org/licenses/by/4.0/}{\includegraphics[width=0.5\textwidth]{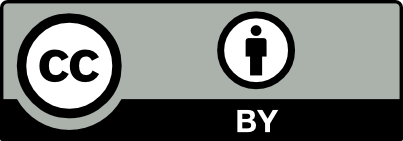}}
\end{minipage}\hfill
\begin{minipage}{0.6\columnwidth}
\end{minipage}\hfill

\href{http://creativecommons.org/licenses/by/4.0/}{This work is licensed under a Creative Commons Attribution International 4.0 License.}}


\renewcommand{\shortauthors}{Ella Dagan, et al.}

\begin{abstract}
We present Project IRL (In Real Life), a suite of five mobile apps we created to explore novel ways of supporting in-person social interactions with augmented reality. In recent years, the tone of public discourse surrounding digital technology has become increasingly critical, and technology's influence on the way people relate to each other has been blamed for making people feel ``alone together,'' diverting their attention from truly engaging with one another when they interact in person. Motivated by this challenge, we focus on an under-explored design space: playful co-located interactions. We evaluated the apps through a deployment study that involved interviews and participant observations with 101 people. We synthesized the results into a series of design guidelines that focus on four themes: (1) \textit{device arrangement} (e.g., are people sharing one phone, or does each person have their own?), (2) \textit{enablers} (e.g., should the activity focus on an object, body part, or pet?), (3) \textit{affordances} of modifying reality (i.e., features of the technology that enhance its potential to encourage various aspects of social interaction), and (4) \textit{co-located play} (i.e., using technology to make in-person play engaging and inviting). We conclude by presenting our design guidelines for future work on embodied social AR. 
\end{abstract}




\begin{teaserfigure}
    \centering
    \includegraphics[width=0.9\textwidth]{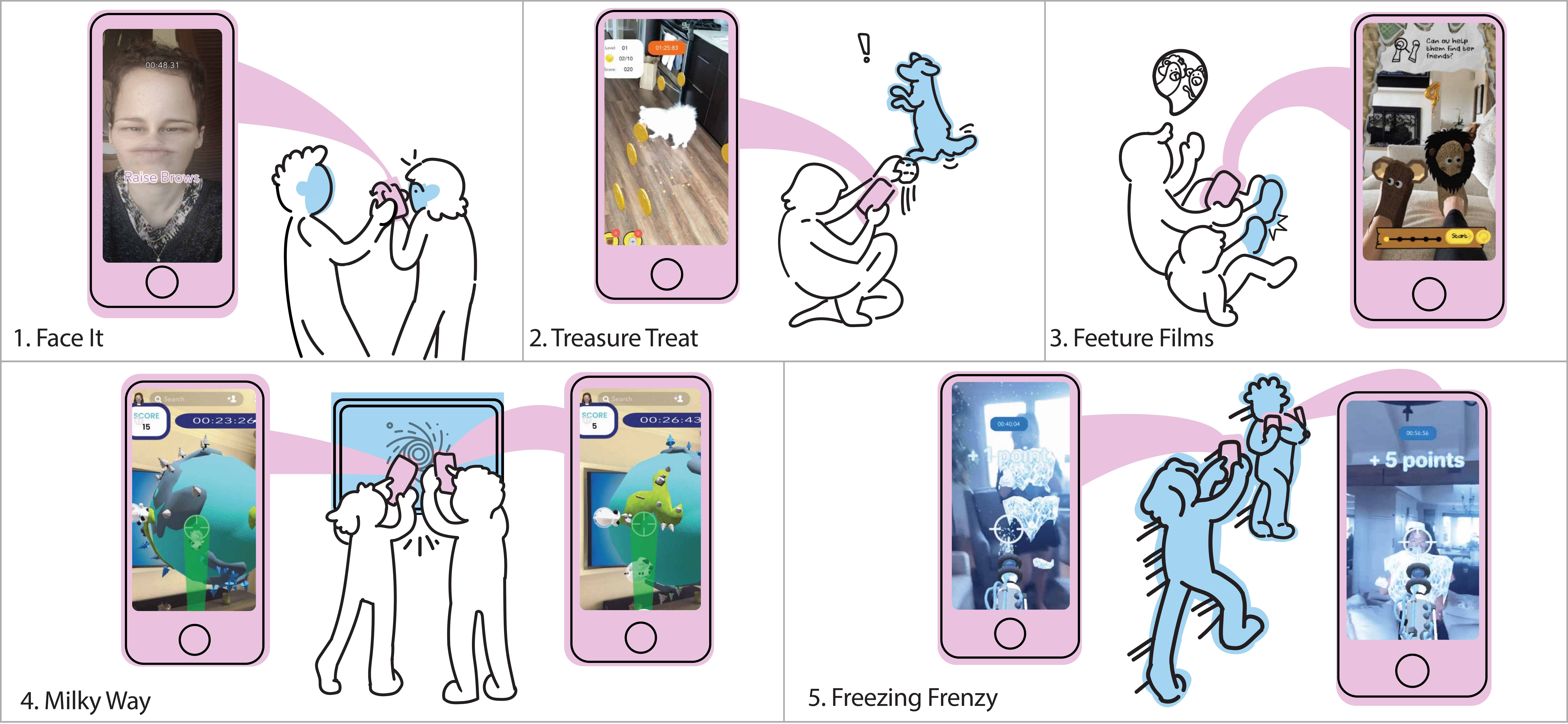}
    \caption{Playful co-located mobile AR apps: (1) ``Face It'' (2); ``Treasure Treat''; (3) ``Feeture Films''; (4) ``Milky Way''; (5) ``Freezing Frenzy.'' Device arrangement is illustrated in pink and enablers in blue.}
    \Description{Five schematic drawings showing players and screenshots of app being used}
    \label{fig:teaser}
\end{teaserfigure}

\maketitle

\section{Introduction}

In recent years, the tone of public and academic discourse surrounding the impact of technology on face-to-face interactions has become increasingly negative. Prior research has shown that although digital technology makes it easier to stay connected, it can also disrupt and alienate people engaged in face-to-face social interactions \cite{oduor2016frustrations, turkle2017alone, misra2016iphone} to the extent that, even when people are physically together, or co-located, they still experience a sense of being ``alone'' in their own digital bubbles \cite{turkle2017alone, rogers2014bursting}. Similarly, journalists have argued that \textit{``human contact is now a luxury good''} and that \textit{``separating from screens is harder for the poor and middle class''} \cite{bowles2019human}. Artists have echoed these sentiments through photographs \cite{REMOVED, Soul-Sucking}, cartoons \cite{rob_2010, boligan_corvo_2013}, and new media \cite{SomebodyAapp}, highlighting the isolating nature of technology. 

Alongside the growing concerns about technology and co-location, human-computer interaction scholars have drawn attention to a gap in the literature on co-located technology \cite{olsson2020technologies}. A similar gap exists in terms of commercial technologies designed for in-person interactions, which are rare, especially outside the productivity domain. For example, none of the ten most downloaded iOS apps are explicitly designed for co-location \cite{apple_inc_2019}. Although the COVID-19 pandemic has increased our reliance on technology for remote interaction, it has also fueled interest in activities for in-person social interactions among people with whom we have close social ties. For example, sales of board games increased during the pandemic \cite{boardgames, classicboard}. 

We took these concerns, together with the literature gap, as an opportunity to reimagine and reinvent ways in which technology can support, rather than detract from, co-located interactions. In a recent study, Liu et al. suggest that \textit{``people enjoy engaging in everyday activities with individuals with whom they have strong social ties''} \cite{CYN}. Along with other designers and researchers (e.g., \cite{ferdous2017celebratory}), we aim to identify opportunities in everyday life where AR can positively influence and provide rich shared experiences.

In this work, we introduce ``Project IRL\footnote{IRL is Internet slang for ``In Real Life''},'' a suite of five mobile apps we created to support playful co-located interactions among friends and family (see Fig. \ref{fig:teaser}), adopting a research through design approach \cite{RtDZimmerman, Gaver2012RTD} in our process. To evaluate the apps, we used interviews and participant observations with 101 participants who played with them in groups of two or more. The five apps use augmented reality (AR) to engage people in the following distinct playful experiences: 

\begin{enumerate}
\item People pass around a phone that instructs them to make faces before time runs out, as silly effects are applied. 
\item Dog owners direct their pets to catch virtual coins spread around the physical space. 
\item Parents and their children engage in a story in which their feet are augmented as sock puppets.
\item People compete to catch as many virtual cows as they can, using their phones to beam them up from a planet on a shared TV screen.
\item People chase each other to cover their opponents with virtual ice before time runs out. 
\end{enumerate}

The core contributions of this work are as follows. First, we developed a set of novel co-located AR systems that effectively enabled people to have fun together. Second, we generated a series of design insights for creating \textit{playful co-located mobile AR experiences} organized around four themes: (1) \textit{device arrangement} (e.g., whether the experience requires people to share a single phone or use multiple phones), (2) the role of \textit{enablers} (i.e., the physical objects that trigger and are the focus of the AR experience), (3) the affordances of \textit{augmentation} (i.e., features of the technology that enhance its potential to encourage various aspects of social interaction), and (4) the \textit{co-located play} experience (i.e., using technology to make in-person play engaging and inviting).
Third, we propose a set of guidelines for designing playful co-located AR experiences (e.g., nudging physical touch, using people's bodies--and even their pets--as enablers, and building on familiar games). We conclude with a call for more research on \textit{embodied social AR}, a generative approach that utilizes enablers to focus on bodies as a means of purposefully supporting in-person social experiences. Our aim is for this work to inspire researchers and technology designers to explore new design avenues for encouraging people to playfully interact in person using mobile AR or other technologies.

\section{Related Work}
\subsection{The need to design for co-location}
\begin{quote} \textit{``Every time you check your phone in company, what you gain is a hit of stimulation, a neurochemical shot, and what you lose is what a friend, teacher, parent, lover, or co-worker just said, meant, felt.''} \\---Sherry Turkle \cite{turkle2016reclaiming} \end{quote} 
Researchers have empirically demonstrated that the mere presence of mobile devices can negatively impact communication during face-to-face conversations \cite{przybylski2013can, misra2016iphone}. Similarly, ethnographic studies have shown that reliance on technology is making people increasingly socially isolated, causing them to experience the sensation of being ``alone together'' when they are in the presence of others \cite{turkle2017alone}. Researchers argue that the human-computer interaction literature requires more research focusing on how to design technology that will encourage co-located interactions by \textit{``not only providing opportunities, but also utilizing computational features that nudge and stimulate people to take action''} \cite{olsson2020technologies}. Olsson et al. \cite{olsson2020technologies} have identified various roles technology can take in human interactions, ranging from \textit{``enabling''} to \textit{``encouraging''} social experiences. They argue that \textit{``future design endeavors would benefit from more deliberate choices of specific phenomena, social settings, target user groups, or type of interaction—particularly those that aim to actively enhance the quality of social interaction''} \cite{olsson2020technologies}. Similarly, Isbister calls on designers to consider the space between people. She argues that we should design interactions that require people to share devices because they can support interdependent interactions \cite{Suprahuman}, i.e., interactions that require people to engage with each other in order to interact with their devices. 

Lundgren et al. \cite{Lundgren2015MobileCollocated} encourage designers to take four perspectives when designing co-located mobile experiences: social, technological, spatial, and temporal. Further, their research suggests that co-located interaction is based on collaboration, communication, competition, or a mix of the three. In our work, we embrace these calls to action by creating a suite of mobile AR applications to encourage co-located socialization. Our exploration considers different interaction types, including competitive and collaborative experiences, and experiments with various social relations, e.g., parents and children, and close friends. 

\subsection{The value of playful casual interactions}
In recent years, researchers have drawn attention to the value of playful interactions \cite{Altarriba2020Play, sharp2019fun}. Therefore, we decided to design applications to enable these types of experiences among people who are together in the same space. In our work, we were interested in creating technologies that support \textit{casual} social interactions, without focusing on productivity (as much of the existing HCI literature have already addressed this, e.g., \cite{lucero2010collaborative, dickey2010lessons, alavi2012ambient, kreitmayer2013unipad}). For this reason, we turned to games and playfulness as the driving force for creating co-located mobile experiences.

In prior work on co-located games, researchers have argued that technology designers should focus on two core elements: game mechanics and the social affordances of the game's interface (i.e., features that encourage various types of social interaction) \cite{Socio-Spatial2008DeKort}. For example, one approach to social affordance is to \textit{``shape the flow of interpersonal distance (proxemics) in pro-social way''} \cite{Isbister2018SocAfford}. More broadly, researchers suggest that game designers who want to enhance socio-emotional player experiences should focus on creating physical-social play \cite{segura2015enabling}. Inspired by prior research on movement-based games \cite{Mueller2014Movement}, we built on two proposed guidelines that relate to AR: \textit{``construct the player's actions in a way that gives room for sensor error without drawing attention to it''} and \textit{``avoid game mechanics that require precise control.''} 

We find the qualities of mobile AR well suited to co-located playful interaction due to its strong support for embodied interaction \cite{embodiedHandHeldAR, dourish2004action} and the fact that AR is grounded in the physical environment \cite{wetzel2008guidelines}. Therefore, we decided to focus on exploring AR as a design material for such experiences.

\subsection{Leveraging Mobile AR in co-location}
\label{sec:lit}
We focused on building our system using AR because, unlike other technologies, it relies heavily on the physical world, i.e., the reality shared by people who are inhabiting the same place at the same time. We surveyed the literature and identified the following five considerations outlining how and why AR technologies are well suited to our task. Research suggests that AR is capable of supporting playful co-located social experiences because it is:

\begin{itemize}

    \item \textbf{Grounded.} AR can cultivate ambiance by stimulating a variety of senses, overlay content onto the real world environment \cite{wetzel2008guidelines}, transform the world into a \textit{``playground''} \cite{Altarriba2020Play}, and support exploration \cite{Paasovaara2017CollocatedPMG}.
    
    \item \textbf{Embodied.} AR can physically and socially facilitate social activities, imbue meaning through embodiment \cite{dourish2004action}, and support novel physical interactions \cite{embodiedHandHeldAR}.
    
    \item \textbf{Playful.} AR can facilitate playfulness through experience design and content that is surprising, humorous, thrilling, or challenging \cite{Altarriba2020Play}.
    
    \item \textbf{Social.} AR can facilitate and enhance various types of social interactions and relationships \cite{wetzel2008guidelines, vella2019sensePMG, BragFish}.
    
    \item \textbf{Memorable.} AR interactions can be memorable (e.g., \cite{koskinen2019player}) due to well-crafted experience design \cite{pine1999experience}; it can also be easily recorded and shared \cite{wetzel2008guidelines}. 
    
\end{itemize}   

Although mobile AR is still an under-explored technology that has not been widely adopted, some inspiring experimental designs support co-located interaction to varying degrees. For example, researchers have explored co-located mobile AR in relation to interfaces for group collaboration \cite{CollabAR2020}, co-creation \cite{guo2019blocks, ARLooper2020}, communication \cite{Raffle_2007}, and shared-world gameplay \cite{Brick2019colocatedAR}. In addition, there are several research projects examining mobile AR game experiences designed for synchronous and co-located interaction, e.g., \cite{Multi-Device2015CaptureFlag, Zambetta2020ARQueues}. Beyond this research, there are a few commercial mobile AR apps that can also be used in co-located interactions \cite{within_unlimited_inc_2020, leverx_2020, paavilainen2017pokemon}. These previous works helped us to explore a different aspect of the co-located mobile AR design space. However, our ultimate goal was to explore how mobile AR can be used as a \textit{design material} to support playful co-located interaction. With this in mind, we we designed five experiences that leverage AR's grounded and embodied qualities.

With this prior work as the backdrop for our research, we explored ways of facilitating playful co-located experiences by experimenting with mobile AR's affordances. Our goal was to provide generative insight. To that end, we took an exploratory and interpretative approach \cite{waern2015experimental} as part of our research through design \cite{RtDZimmerman, Gaver2012RTD} process. 

\section{Project IRL: A suite of five Mobile Augmented Reality apps}
\begin{table}[th]
\centering
\footnotesize
\renewcommand{\arraystretch}{1.3}
\caption{AR considerations and operationalized design attributes.}
\label{fig:IRL-Tables-ARConsiderations}
\centering

\begin{tabular}{p{0.1\textwidth}p{0.08\textwidth}p{0.18\textwidth}p{0.12\textwidth}p{0.14\textwidth}p{0.15\textwidth}}
\toprule
\multicolumn{6}{c}{\bfseries Mobile AR Considerations}\\\cmidrule{2-6}
&\multicolumn{1}{c}{\textbf{Grounded}} & \multicolumn{1}{c}{\textbf{Embodied}} & \multicolumn{1}{c}{\textbf{Playful}} & \multicolumn{1}{c}{\textbf{Social}} & \multicolumn{1}{c}{\textbf{Memorable}} \\\cmidrule{2-6}
\hspace{2mm}\textbf{Design}\enskip \textbf{Attributes} & \enskip Enabler & \mbox{Device Arrangement} \enskip Physical Movement & Augmentation & Interaction Type & Camera Recording\\
\bottomrule
\end{tabular}
\end{table}

\subsection{Design Process}
During our research through design \cite{RtDZimmerman, Gaver2012RTD} process, we asked ourselves two questions: \textit{``what makes this design work only when people are together in person?''} and \textit{``what would make this design enjoyable or meaningful to use when interacting with others in person?''} This approach proved generative and kept our brainstorming sessions focused on co-located social experiences. We drew inspiration from other researchers' proposed guidelines for designing AR games that emphasized \textit{``experience first, technology second''} \cite{wetzel2008guidelines}. Further, we built on interactions that are already familiar to people by finding inspiration in existing games (e.g., ``Fetch''\cite{FecthGame} or ``Tag''\cite{wise_forrest_2003}) and social activities (e.g., storytelling). We took the following steps (over a three-month period): 

\begin{enumerate}
  \item \textit{Formulating design attributes.} We identified five design considerations for mobile AR to support co-located interactions based on our literature review (section 2.3). We operationalized these design considerations into concrete techniques by mapping them to \textit{design attributes} (Table \ref{fig:IRL-Tables-ARConsiderations}). Each has a range of possibilities; for example, an enabler could be a body part, a pet, or a physical object like a screen; interaction types could be competitive or collaborative, etc.
 
  \item \textit{Generating ideas to prototype.} Based on these techniques, our research team (which included designers, engineers, and researcher-focused team members) brainstormed both individually and collectively in remote video sessions, sharing our design ideas via Google slides. We generated a list of over twenty app ideas with various combinations of the target design attributes.
  \item \textit{Prototyping selected concepts.} We selected five concepts from the list of ideas to develop as apps. When selecting, we tried to have the broadest coverage of the matrix of combinations of the design attributes (step 1) while balancing the practical constraint of available time. When we were prototyping the concepts, we frequently met to discuss design iterations and internally play-test new versions. The design attributes for each of the five IRL apps are specified in Table \ref{fig:IRL-Tables-Attributes}.

\end{enumerate}

We designed the augmentation in our apps to be the driving force of the experience and tied it to \textit{enablers}. Enablers are physical entities that trigger--and are the focus of--the AR experience. Enablers are similar to markers \cite{fiala2005artag, hirzer2008marker} in triggering AR objects, but they differ in that they can be any physical entity (beyond a 2D pattern), and they play an integral role in the AR experience. We chose to experiment with distinct enablers, such as pets, parts of people's bodies (e.g., faces, feet, or entire bodies), and the use of shared screens (e.g., TVs).

\begin{table}[t]
\footnotesize
\renewcommand{\arraystretch}{1.3}
\caption{Design attributes for each of the AR apps.}
\label{fig:IRL-Tables-Attributes}
\centering

\begin{tabular}{p{0.145\textwidth}p{0.135\textwidth}p{0.11\textwidth}p{0.17\textwidth}p{0.11\textwidth}p{0.18\textwidth}}
\toprule
\multicolumn{6}{c}{\bfseries Design Attributes}\\\cmidrule{2-6}
\textbf{IRL App} &\textbf{Device \mbox{Arrangement}} & \textbf{Enabler} & \textbf{Augmentation} & \textbf{Interaction Type} & \textbf{Physical Movement} \\
\midrule
\textbf{Face It} & One phone & Face & \mbox{Face filters} (i.e., "lenses") & Competition & Gesturing with face; passing an object\\
\textbf{Feeture Films} & One phone & Feet & Sock puppets & Storytelling & Gesturing with feet\\
\textbf{Treasure Treat} & One phone & Dog & \mbox{Dog's silhouette and} \mbox{environment coins} & Cooperative Game & Gesturing to the dog; dog moving around \\
\textbf{Milky Way} & Two phones, one TV or laptop & Video on TV & 3D planet & Competition & Swiveling around the screen \\
\textbf{Freezing Frenzy} & Two phones & Full body & Ice on body  & Competition & Chasing other players \\
\bottomrule
\end{tabular}
\end{table}

\subsection{App Descriptions}

Here we describe each of the five apps\footnote{See video: \href{https://youtu.be/\_51-IlxdBjg}{https://youtu.be/\_51-IlxdBjg} and project website: \href{https://letsplayirl.com}{letsplayirl.com}}\footnote{You can find access codes to download the apps using Snapchat on Android or iOS in the supplemental materials}, explain the user experience of each, and highlight (in bold text) all of their associated design attributes (See Fig. \ref{fig:teaser} and Table \ref{fig:IRL-Tables-Attributes} for an overview). The apps represent a variety of activities with several different contexts of use.

\subsubsection{Face It (FI)}

\begin{figure}[h] \centering \includegraphics[width=0.8\linewidth]{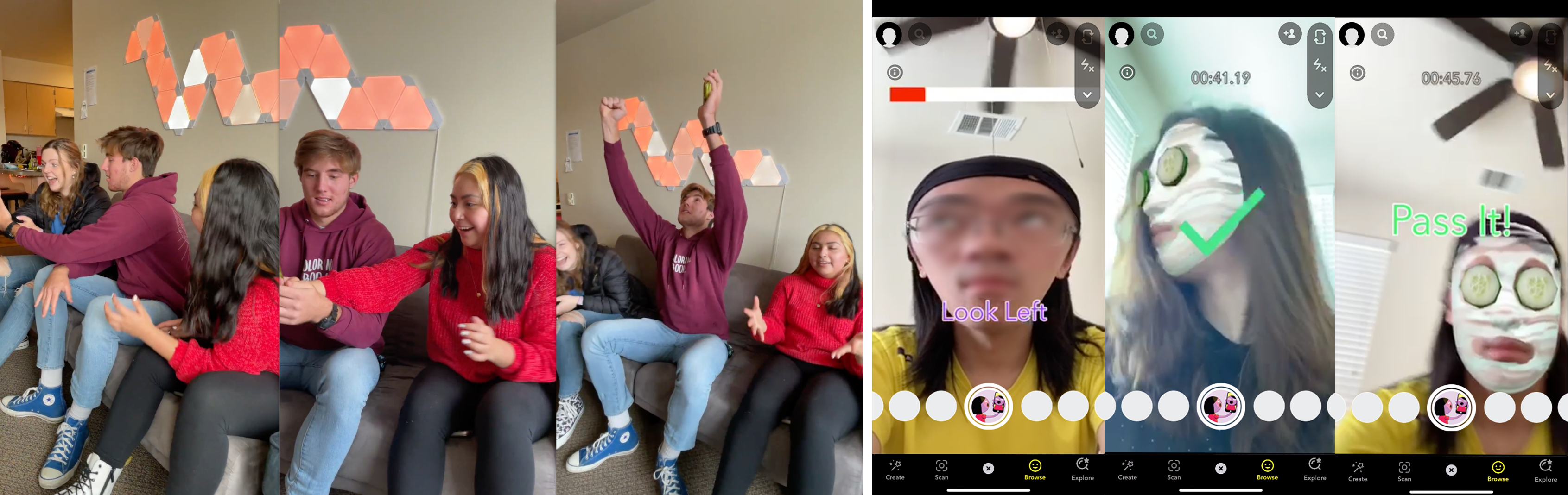} \caption{\label{fig:FI-Mixed} Players passing the phone in a play session with the FI app (left). Snapshots of the FI app instructing players to make a facial gesture and pass the phone (right). Note that we blurred the faces of study participants to preserve their privacy. Images not blurred are people hired to pose for videos.} \end{figure}

We were inspired by the game mechanic of \textit{passing an object between people}, an interaction many people are familiar with from toys like "Bop-It" \cite{BopIt} or games like "Hot Potato" \cite{wise_forrest_2003}. We were also inspired by popular face lenses on Snapchat \cite{snap_face_filter} and filters on Instagram \cite{Instagram_filters}, particularly, a co-located game that uses head tilt as the enabler \cite{lim_2020}.

To play FI, multiple players \textbf{pass one phone around to compete with each other}. FI is enabled by using the phone's front-facing camera to track and augment \textbf{players' facial gestures}, and the app detects the facial expressions of one person at a time, turning the \textbf{face} into the game's controller (i.e., players' faces enable the AR experience). Players see their faces augmented continuously in humorous ways through face distortions. For example, their face turns into an animated broccoli, among other \textbf{face filters} (see Fig. \ref{fig:FI-Mixed}).

The app asks the players to begin the game by pressing a video recording button that starts a 60-second timer. FI prompts players to perform different facial gestures through text and audio instructions, e.g., ``smile,'' ``kiss,'' and ``look left.'' Unexpectedly, the app shows and plays the prompt ``pass it'' to ask players to pass the phone quickly to the next player, who then receives their own set of prompts. We developed FI using Lens Studio face segmentation and classification models for \textit{facial gestures}. More specifically, we used Lens Studio's face triggers \cite{LensStudioBehavior} to determine if the players completed their prompts before their time ran out. 

The pace of the music and instructions increases over time, and the amount of time players have to complete each gesture decreases, making it progressively more challenging. When a player fails to execute their instructions, the game round ends, and that player is out. The remaining players continue until only one player is left, and this last player is declared the winner. At the end of each round, players can view a short \textbf{recorded video} of the gameplay that they can save to their phones or share with others. 

\subsubsection{Feeture Films (FF)}
Several research projects explored augmenting traditional storytelling with digital technology. For example, researchers developed a room-sized mixed reality immersive storytelling experience for children to experiment with \cite{alborzi2000designing}. Others explored a storytelling system to foster creativity and collaboration among children by mixing physical and digital story elements \cite{cao2010telling}. Researchers also explored a storytelling AR system with finger puppets \textit{``to enhance social pretend play''} \cite{FingARPuppet}. There are also commercial AR storytelling apps for children (e.g., \cite{within_unlimited_inc_2020}).

\begin{figure}[t] \centering \includegraphics[width=0.8\linewidth]{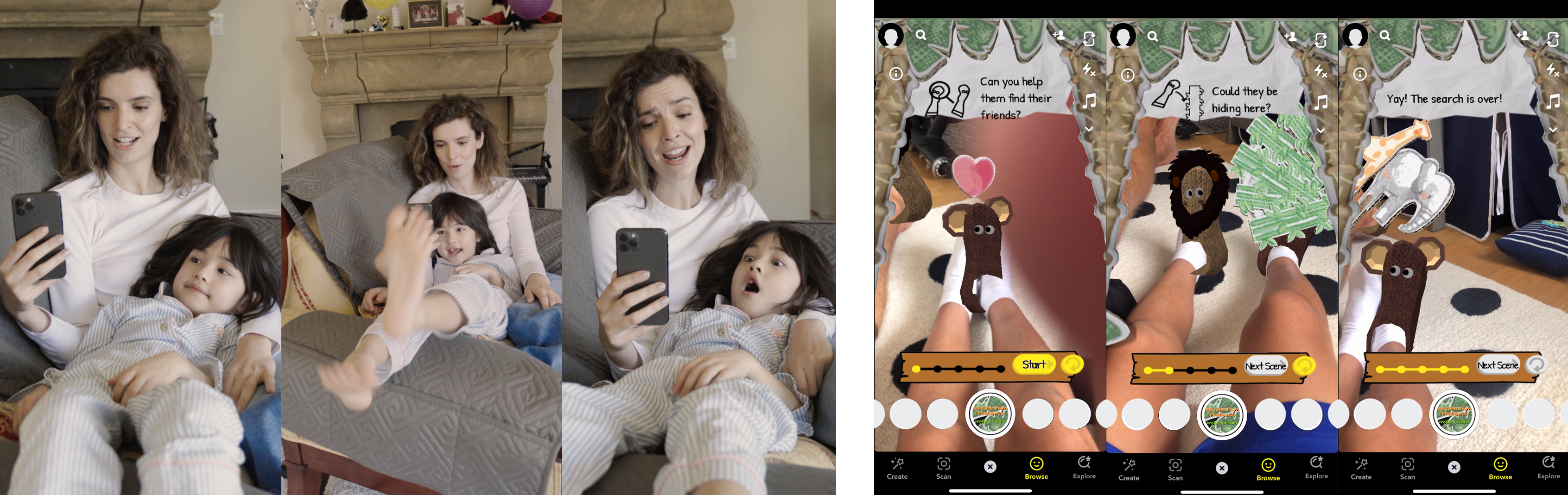} \caption{\label{fig:FF-Mixed} An adult and child point a phone at their feet to augment them using the FF app (left). Snapshots of the FF app showing players' feet augmented as sock puppets (right).} \end{figure}

All of these prior works, along with children's flap books \cite{russo_2020}, inspired us to create FF as an interactive AR \textbf{storytelling experience} for children and parents to share. To use FF, a parent and a child \textbf{share one phone}. The experience is \textbf{enabled by their feet}: the app detects players' feet, using them as the target of the augmentation by taking advantage of Snap Machine Learning's Foot Tracking, which allowed us to detect players' \textit{right and left feet} and estimate their 3D position. \cite{snap_foot_tracking}) (see Fig. \ref{fig:FF-Mixed}).  

FF prompts the parent and child to sit next to each other, side by side (typically on a couch or bed), and uses one foot from each of them. Their \textbf{feet are augmented as sock puppets} that go on a jungle-themed adventure together to find their friends. \textbf{Foot gestures} are used to interact and trigger the AR story-objects; for example, players hover on overlaid objects to discover other animals or tap their feet together to reveal their emotions (which are displayed with emoticons).

FF has four scenes. The parent and child can navigate and experiment with verbally adding their own stories or enact the story by moving their sock puppets in the frame. In addition, they can capture parts of the experience by taking photos or videos; they can also exit the app and return to where they left off in the story at a later time (we utilized internal storage to store the scene ID). FF is a unique form of digital storytelling as it encourages experimentation with body gestures while supporting in-person interaction between a parent and a child. We designed it to keep the traditional storytelling interaction of reading stories aloud while injecting bodily play and experimentation with digital augmentation. 

\subsubsection{Treasure Treat (TT)}
\begin{figure}[b] \centering \includegraphics[width=0.8\linewidth]{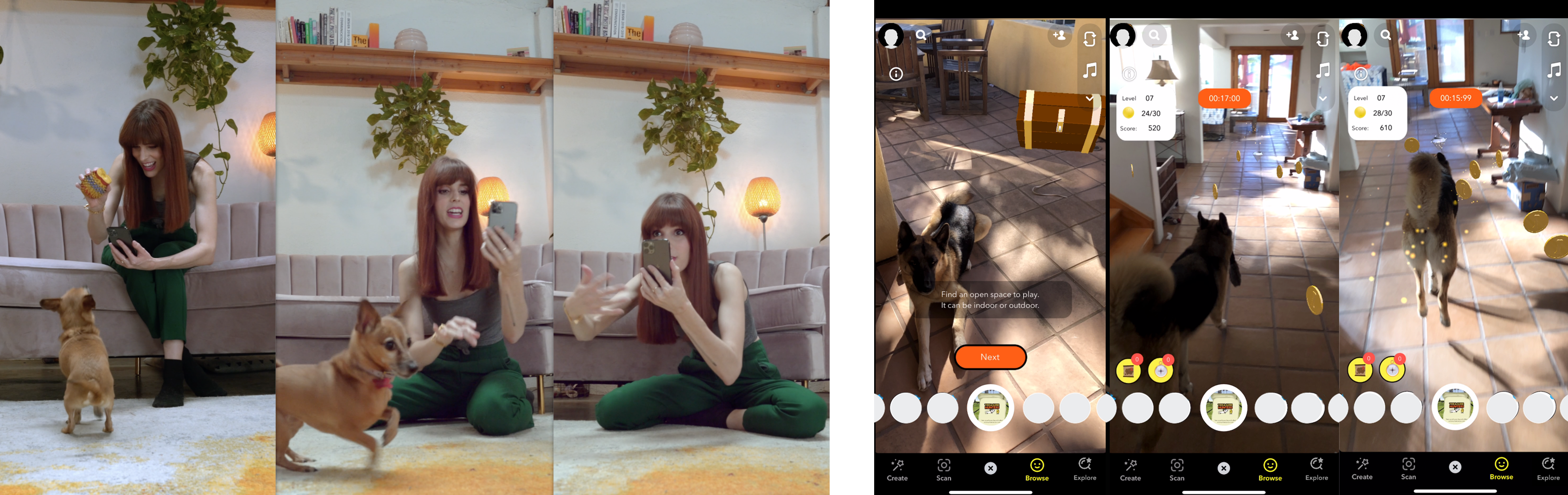} \caption{\label{fig:TT-Mixed} A pet owner and their dog using the TT app (left). Snapshots of the TT app showing the dog collecting points by colliding with AR coins (right).} \end{figure}
This app explores playful co-located interaction between a person and their dog using mobile AR. Human-pet relationships are an essential part of the co-located interaction design space. Further, due to COVID-19, people are spending more time at home and adopting dogs at higher rates \cite{morgan2020human} than before the pandemic. AR technology has started to follow this pet trend too. For example, GoDog \cite{leverx_2020} is an AR dog training app, and Lens Studio has added an API \cite{snap_inc_pets} that lets developers create AR apps that use dogs' and cats' faces as enablers. More broadly, we found inspiration in existing games people play with their dogs (e.g., "Fetch" \cite{FecthGame}) as a way of including physical movement. 

TT is a pirate-themed \textbf{cooperative game} in which the player and their dog team up to collect \textbf{AR coins} that appear from a treasure box and then scatter on the floor. TT is a level-based progression game, which is also inspired by classic video games such as Super Mario \cite{nintendo_smb}. \textbf{The dog's body} enables the game when it is detected (using the phone's rear camera). Then TT instructs the person to tap on the screen to identify the floor's plane. The \textbf{dog's silhouette is also augmented}, and it controls the game: collision detection evaluates whether the coins and the dog's body appear to touch. When the dog is successfully detected, the coins are collected, and the dog-human team earns points. We applied a dog machine learning model to the camera input to detect the dog's position in the camera frame (see Fig. \ref{fig:TT-Mixed}) and to allow us to attach a boundary animation to the dog's body.
 
The human player is expected to motivate and guide their dog partner to move in specific directions in order to collect all of the available coins within a time limit. People might do this by, for example, \textbf{gesturing to the dog to move} in whichever way they want, making sounds to attract the dog's attention, pointing in a specific direction (e.g., \cite{kaminski2013dogs}), throwing toys, and using treats. Each level introduces a new challenge by changing the number of coins or the time constraints. If the team successfully finishes collecting all of the coins on time, they proceed to the next level. If not, the game is over. If they complete all of the challenges, they win a "pirate" AR filter for the dog.

\subsubsection{Milky Way (MW)}
Inspired by the recent concept of ``Augmented Reality Television'' (ARTV) \cite{vatavu2020conceptualizing}, we explored the use of a second screen (TV or laptop) for a multiplayer game design that \textit{``employs one main screen shared by two players, each one also using a second (private) screen''} \cite{pagno2015guidelines}). 

We drew from the familiar co-located interaction of gathering around a shared focal point to play (e.g., playing video games with others). MW is a \textbf{competitive game} and is an intergalactic AR twist on the game "Whack-a-Mole" \cite{Wack-A-Mole}. Two or more players are expected to gather around a TV or laptop to scan a YouTube video using the rear cameras on their \textbf{individual phones} to augment a second screen. The \textbf{YouTube video} enables the AR experience and allows the game to sync automatically. The screen display serves as a physical anchor for \textbf{players to swivel around}. To find the video on YouTube, players need to search for a five-letter keyword.

The video shows a rotating "black hole." We used the angle of rotation to synchronize all of the player's devices without the need for a network or Bluetooth connectivity between the devices. The rotation itself becomes a unique marker that allows the system to determine when to start the game and make new objects appear on the screen. This enabled players to begin the game simultaneously without the need for network connectivity. Further, this approach made tracking consistent between phones, regardless of the time or angle at which the marker is scanned. Finally, we used \textit{2D markers} to track and augment objects on a video.  

\begin{figure}[b] \centering \includegraphics[width=0.8\linewidth]{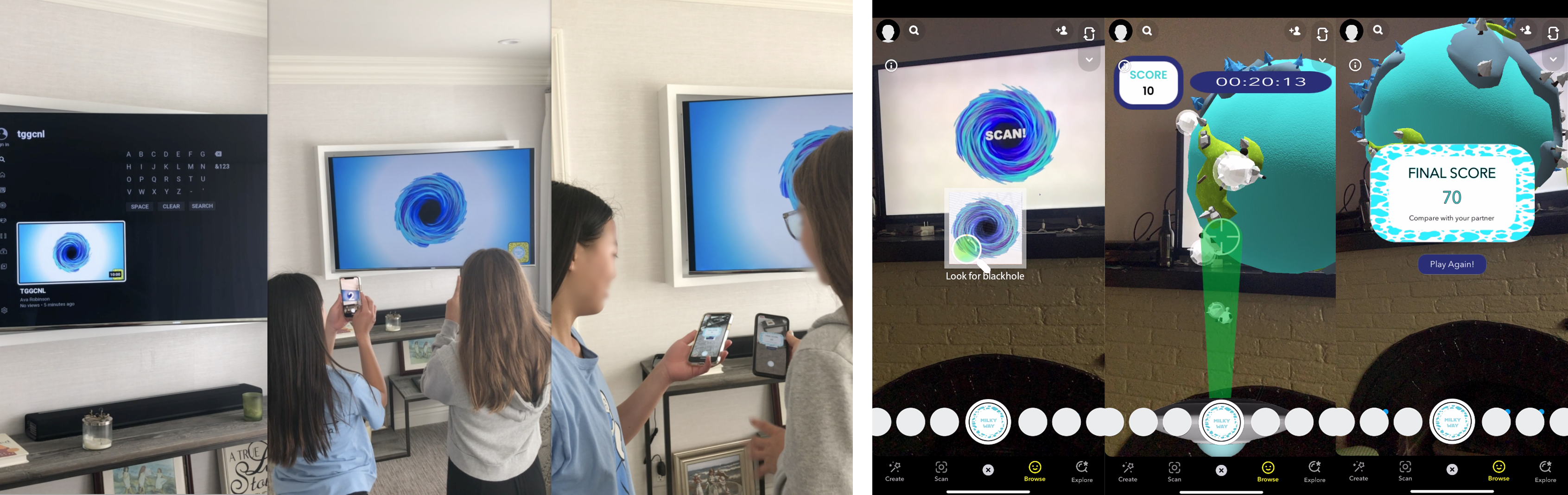} \caption{\label{fig:MW-Mixed} Players using the MW app pointing their phones at a TV (left). Snapshots of the MW app showing the rotating marker and the AR planet where cows must be rescued (right).} \end{figure}

MW starts when a big \textbf{3D planet} inhabited by space cows appears on players' mobile screens, as if they are coming out of the video. The goal is to abduct as many space cows from the 3D planet as possible (see Fig. \ref{fig:MW-Mixed}). When MW starts, it generates a seed from the start timestamp on the players' phones. This seed randomizes the space cows' appearance in the game sessions (but maintains consistency during the session) such that players see the space cows appear at the same positions and times on the planet. Players can move around the 3D planet, aim, and then tap the screen to abduct the space cows during the 30 seconds of game play. Players see their scores and receive a prompt to compare them to determine the winner.

\subsubsection{Freezing Frenzy (FR)}

\begin{figure} \centering \includegraphics[width=0.8\linewidth]{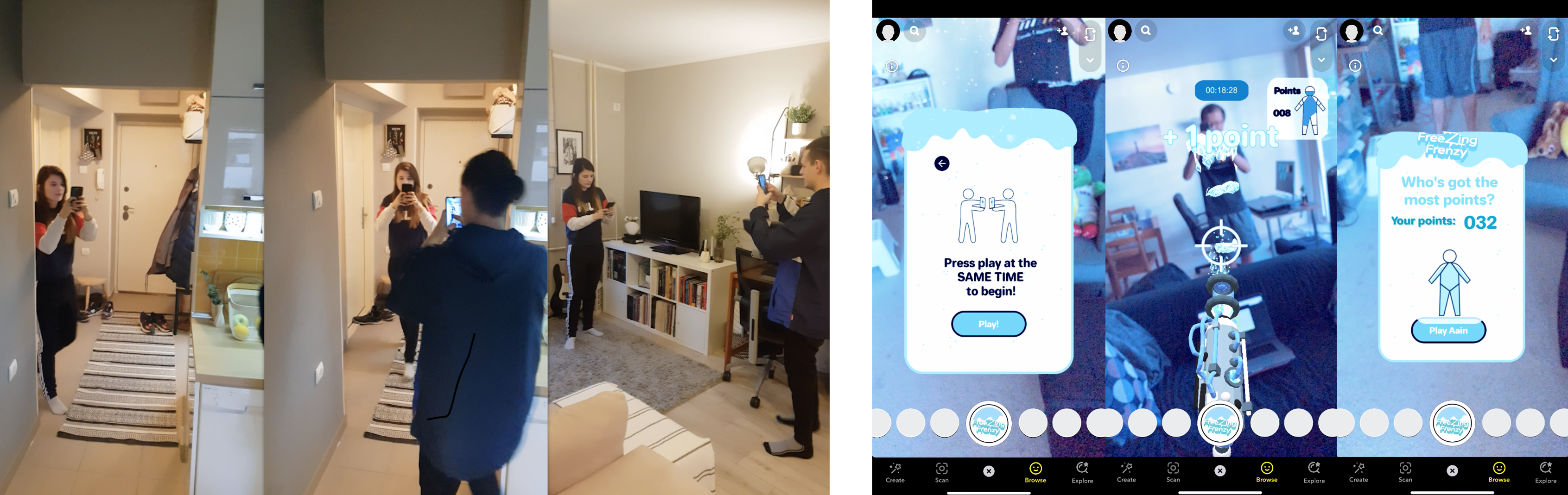} \caption{\label{fig:FR-Mixed} Players using the FR app and pointing their phones at each other (left). Snapshots of the FR app showing the frozen AR effect over players' bodies when they get hit (right).} \end{figure}

FR is an AR battle \textbf{competitive game} inspired by the classic "Laser Tag" \cite{ferret_2007} game. With this app, we focused on using players' \textbf{entire bodies} as enablers, both to trigger the game and to serve as the target of the augmentation. We also draw ideas from physical AR battle interactions, such as AR pong battle \cite{Hu_2018} and a multiplayer AR shooter game \cite{hollister_2017}.

FR allows two players to use their \textbf{phones in parallel} to simultaneously target each other's entire bodies with their rear cameras. For a 30-second interval, the phones are "transformed" into freezing guns. Players \textbf{chase each other} and tap on the screen to shoot and "cover" their opponent's body with \textbf{AR ice}. We chose the "freezing" theme because attaching ice texture to the body rendered relatively well. If players make a successful "hit," chunks of AR ice appear to cover their opponent's body and they are awarded points (see Fig. \ref{fig:FR-Mixed}).

FR prompts players to manually coordinate the start of the game by pressing a button simultaneously, and it prompts them again at the end to compare their scores. We developed this app using Lens Studio's "Full Body Attachments" \cite{LensStudioFullBodyAttachments} to detect bodies and distinguish between their body parts (e.g., head, neck, and forearm). This feature relies on 2D body segmentation models. We also used it to detect target collisions between a tap on the screen and the player's body to initiate AR ice shooting.

\section{Evaluation}

The COVID-19 pandemic required us to adapt our plan to support remote evaluation. Our study design consisted of one-hour video call sessions. Our data collection included (i) observational notes on participants using the apps, (ii) transcribed semi-structured interviews, and (iii) participants' screen recordings. We based our semi-structured interview questions on recommendations developed by Fullerton and colleagues \cite{fullerton2014game}, and additional open questions about the overall experience of using the apps with others. We ran four pilot sessions with participants, then implemented small changes in our protocol before deploying the study. To study mobile AR apps intended to support playful co-located interaction, we created participant groups of two co-located people (with the exception of one group of three).

\subsection{Study Protocol}
We presented participants with a short slideshow introducing the study, the apps, and the planned activities for the hour and asked them to confirm consent to record. We then sent a link to a short demo video so that they could watch others using the app. Finally, we gave them an access code to use the app and asked them to start screen recording on their phones before using it. We instructed participants to use the app as much as they wanted (while we kept track of the time, stopping them only if they took over 20 mins). While they used it, we turned off our video and audio on the call to minimize interruptions during their play experience. When participants finished playing, we asked for their feedback in a semi-structured interview (approx 20 mins). Finally, we asked them to share their screen recordings by uploading them to a shared drive.

\subsection{Participants}
We recruited participants (all unique subjects) by posting on Slack channels at a technology company, on Facebook groups, and on university mailing lists (see Table \ref{tab:participants-app}). Participants were offered gift cards to compensate them for their time\footnote{We asked for participants' consent to use their faces and photos in our research and in future publications.}. 
\begin{table}[th]
\footnotesize
\renewcommand{\arraystretch}{1.3}
\caption{Groups of study participants per app. TT: ten dogs considered participants; FF: One adult and one child per group.}
\label{tab:participants-app}
\centering
\begin{tabular}{{p{0.2\textwidth}c@{\hspace{8mm}}p{0.40\textwidth}}}
\toprule
\textbf{App} & \textbf{Groups} & \textbf{Participants} \\
\midrule
Face It (FI) & 10 & n=21; Age range 14-55
\\
\hline
Feeture Films (FF) & 10
& n=20 (10 children, 10 adults); Age ranges 4-7 \& 30-45 \\
\hline
Treasure Treat (TT) & 10 & n=20 (10 dogs, 10 people); Age range 24-43 \\
\hline
Milky Way (MW) & 10
& n=20; Age range 19-61\\
\hline
Freezing Frenzy (FR) & 10 & n=20; Age range 20-39 \\
\bottomrule
\end{tabular}
\end{table}

\subsection{Analysis}
We used a hybrid qualitative method of thematic analysis \cite{joffe2012thematic}, including deductive and inductive approaches. To analyze our participant data, we examined how participants used each app in relation to the five AR considerations (that we identified from the literature in section 2.3) to help us articulate why mobile AR is effective for co-located interaction. As mentioned earlier, our data collection included interview transcripts and participants' observations. 

Three researchers conducted a bottom-up thematic analysis in codebooks on the interview transcripts to analyze the data. The three researchers developed a codebook template to organize text for subsequent interpretation. The codebooks had initial categories relating to the five AR considerations (grounded, embodied, playful, social, and memorable). For each app, the researchers discussed and adjusted the codes based on their specific design attributes. Independently, the researchers performed open coding on the transcriptions of the audio-recorded interviews. They then met to discuss and iterate the codebook based on similarities in participants' data while also taking note of unique responses. Researchers also added their observational notes to the codebook, using participants as the core unit of analysis.

\section{Results}
\label{sec:results}
Mentions of the design attributes (see Table \ref{fig:IRL-Tables-Attributes}) were often evident in participant responses, and their outcomes were observed in their behaviors. Below we synthesize the results under four main themes that relate to the design attributes: (i) device arrangement: one device per group, one device per player; (ii) the roles of enablers; (iii) affordances of augmentation; (iv) co-located play. For each, we created subcategories to articulate how we could use mobile AR to create co-located play experiences (see Table \ref{fig:IRL-Tables-Results}).

\begin{table}[b]
\small
\renewcommand{\arraystretch}{1.3}
\caption{Results: four main themes and subcategories.}
\label{fig:IRL-Tables-Results}
\centering
\footnotesize
\setlength{\tabcolsep}{3pt}
\begin{tabular}{p{0.18\textwidth}p{0.2\textwidth}p{0.27\textwidth}p{0.29\textwidth}}
\toprule
\textbf{Device Arrangement} &\textbf{The Roles of Enablers} & \textbf{Affordances of Augmentation} & \textbf{Co-located Play}  \\
\midrule\\[-16pt]
\hspace{-6mm}\parbox[t]{0.21\textwidth}{
\begin{itemize}
    \item Shapes proxemics\vspace{3pt}
    \item Coordination (\mbox{which nudges} communication)
\end{itemize}}&
\hspace{-6mm}\parbox[t]{0.22\textwidth}{
\begin{itemize}
    \item Social focal point\vspace{3pt}
    \item Moving together
\end{itemize}}
&
\hspace{-6mm}\parbox[t]{0.27\textwidth}{
\begin{itemize}
    \item Entertaining to watch\vspace{3pt}
    \item Experimenting together
\end{itemize}}
&
\hspace{-6mm}\parbox[t]{0.33\textwidth}{
\begin{itemize}
    \item Enhancing friendly competition\vspace{3pt}
    \item Room for more\vspace{3pt}
    \item Interactive screen time
\end{itemize}}\\\bottomrule
\end{tabular}
\end{table}

\subsection{Device Arrangement: One Device Per Group, One Device Per Player}
\subsubsection{Device arrangement shapes proxemics (27 participants).}
\label{sec:proxemics}
\begin{figure}[] \centering \includegraphics[width=0.7\linewidth]{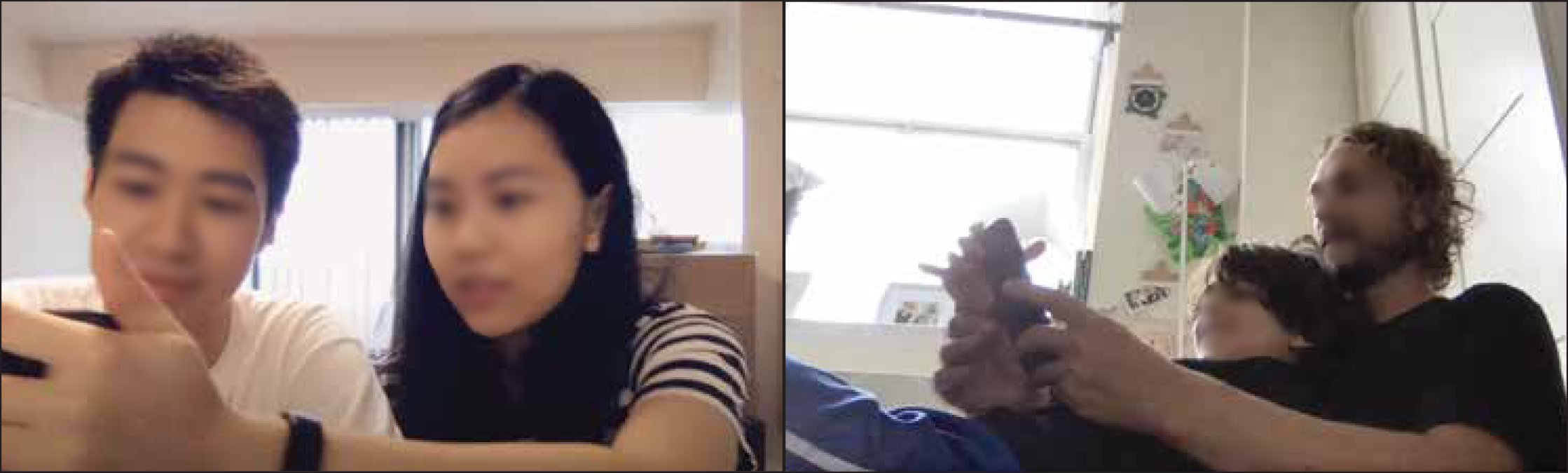} \caption{\label{fig:balancing_distance} Left: FI participants were physically close to each other in order to pass their shared phone. Right: FF participants sat close together in order to share one phone.} \end{figure}

The device arrangement relates to the \textit{shaping of proxemics} social affordance, defined as making 
\textit{``use of sensors to shape the flow of interpersonal distance [...] in pro-social ways}, which \textit{"guides mutual attention through strategic use of feedback to players''} \cite{Isbister2018SocAfford}. In our case, some of the apps required that participants physically moved closer to each other in order to play, so the device arrangement creates changes to social proxemics \cite{marquardt2015proxemic}. For example, Feature Films participants appreciated the fact that playing the game required them to be in close proximity to others-- a participant noted that they felt it was special that the app directed them to sit \textit{``side by side''} with their child (FF-P9) (see Fig.
\ref{fig:balancing_distance}). On top of that, using a shared device made participants collaborate to play and sometimes engage in social touch. For example, FF-P3 said \textit{``So he was sitting down, I was doing the voices, and he was holding the camera on his own feet [...], I added my own character, I added my little finger guy in there and tickled him a little [laughing]''}.

In another case, participants also indicated that there were some challenges associated with needing to be physically close: when playing Milky Way, participants sometimes felt they had to compete for space (e.g., \textit{``the fact that we were using the same laptop. We're in each other's way a little bit. But, [this] also seems like part of the fun of it''} (MW-P19).

\subsubsection{Coordination nudges communication (20 participants).} \label{sec:coordination} 
When the arrangement of devices relies on multiple players using one phone, this creates a situation in which players may need to coordinate their bodily movements. Often, this requires them to find ways to communicate in order to do so. For example, when we observed participants playing Face It they communicated to coordinate transferring the phone between them. One noted that sharing and passing around one phone \textit{``encourage[d] more face-like interpersonal interaction''} (FI-P21), and a Feature Films participant noted that coordinating their movement was \textit{``more of a communication challenge''} (FF-P7). We also observed Feature Films participants communicating to adjust their bodies to fit their feet in the frame and discussing who would hold the phone; they also needed to communicate to plan the coordination of their foot movements to control the app.

\subsection{The Roles of Enablers}
\subsubsection{Enablers can become the social focal point (35 participants).} \label{sec:bodies}

 \begin{figure}[t] \centering \includegraphics[width=0.6\linewidth]{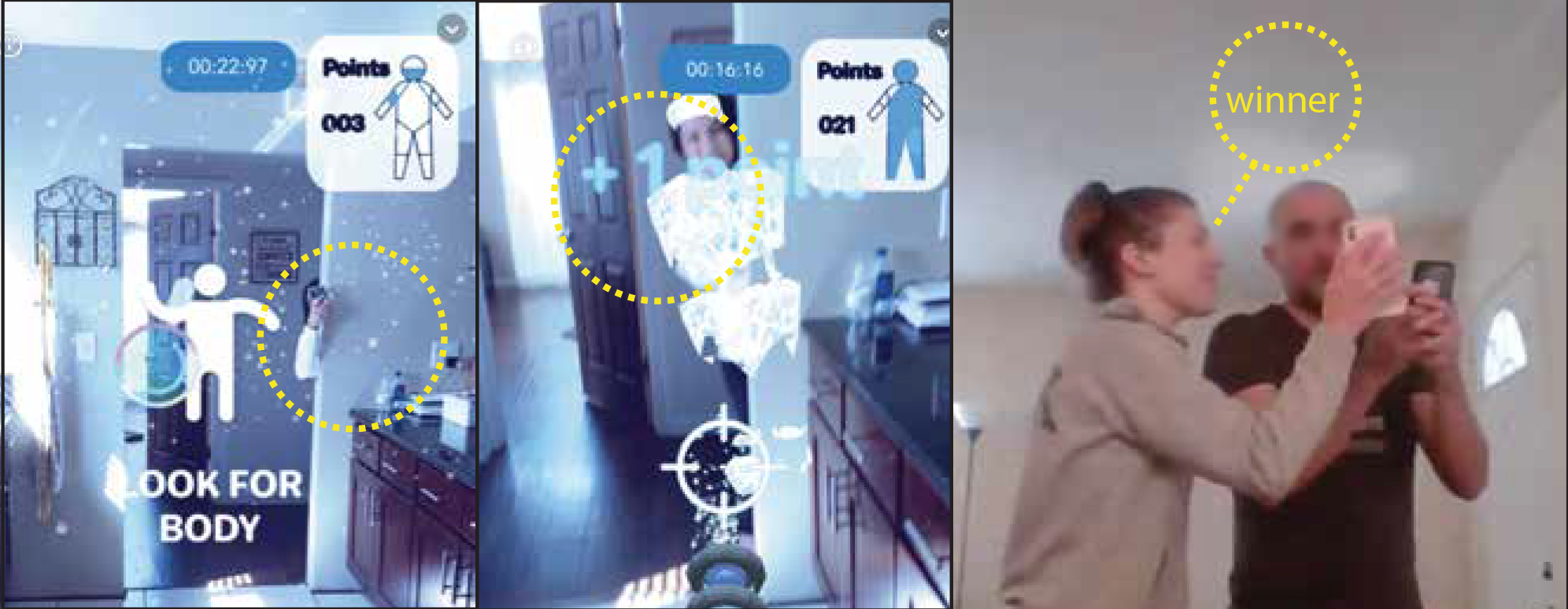} \caption{\label{fig:social_focal_competition} Left: FF Screen shots show how participants focused their phone's cameras at each other. Right: FF players compare their scores.} \end{figure}
 
An enabler is a physical entity that triggers and is the focus of the AR experience. In each of the five apps, we explored a different enabler. In our synthesis, we noticed that specific enablers were better at drawing participant focus toward other players. In particular, enablers that used other players' bodies or body parts were well equipped to facilitate this social focus.

For example, the game mechanics of Freezing Frenzy encouraged participants to focus on each other (by using their own phone to augment the other player's body, see Fig.
\ref{fig:social_focal_competition}), e.g., one participant observed \textit{``I was just so focused on freezing him''} (FR-P6). Freezing Frenzy participant comments indicated that they liked this aspect of the app, e.g., \textit{``I like that I can actually see my brother on the screen''} (FR-P7). In Treasure Treat, players direct the camera on their phone to track their dog's movement. One commented on how they felt synced: \textit{``he looked at me and the [...] speed with which he let me have the toy back--it just really gave me the feeling that he knew that we were working on something [...]. It felt like he knew that we were synchronized on getting something done''} (TT-P2). Another participant observed: \textit{``she saw my excitement and I think she was excited by my excitement''} (TT-P5). Participants also indicated that playing Face It was enjoyable because it included watching others play, e.g., one noted that they would \textit{``rather play it with other people because it's kind of fun to watch them do it, too''} (FI-P11).

\subsubsection{Enablers can encourage moving together (40 participants).} \label{sec:move} 
Enablers are at the heart of the AR experience and afford a wide range of bodily play. Depending on their design, they can support players moving together. 

\begin{figure}[] \centering \includegraphics[width=0.7\linewidth]{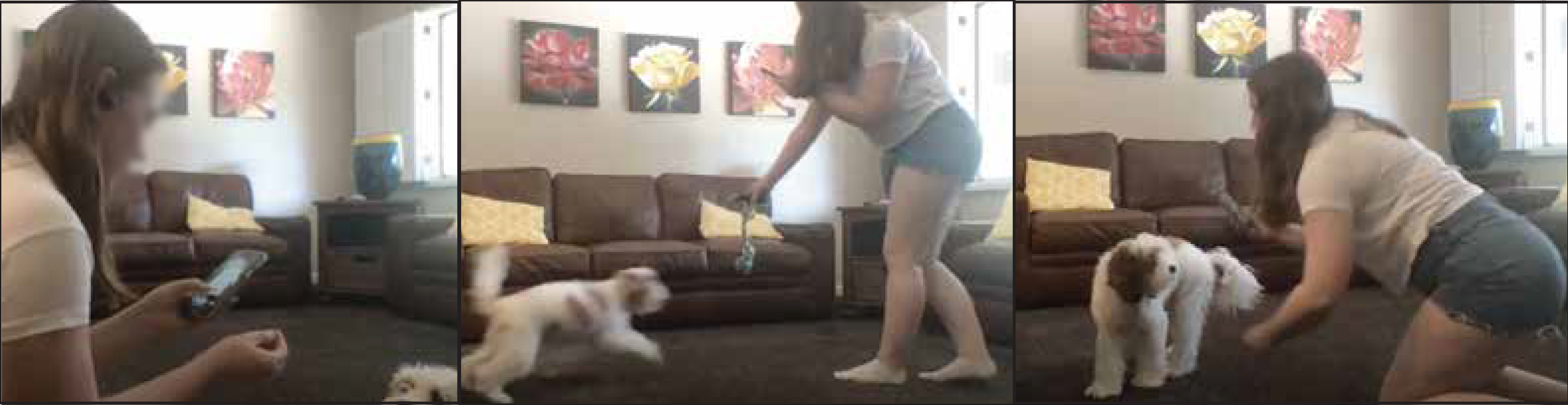} \caption{\label{fig:moving_together} TT participants lured their dogs to play and move around by using treats, toys, and commands.
 } \end{figure}

For example, having participants interact by tapping their feet in Feature Films encouraged social touch and movement. Participants appreciated this aspect of the app, e.g., one noted that \textit{``the cool thing about it was being able to know when you're tapping [your] feet together to get those reactions''} (FF-P1). One participant also commented on the associated intimacy of such interaction, e.g., \textit{``putting your feet together [is something you do] with someone you're really close to''} (FF-P16). In Freezing Frenzy, detecting another player's body enabled the augmentation. Participants realized that they \textit{``should move [their] body so that it can be more exciting''} (FR-P4). Participants noted that this got them \textit{``moving and interacting with [their] friends''} (FR-P13), they appreciated \textit{``that it got [them] up out of [their] chairs''} (FR-P9), and noticed they were \textit{``actually moving around for 30 seconds''} (FR-P9). Similarly, in Treasure Treat, a dog's silhouette enabled the AR experience while its collision with AR coins was at the center of the game. Participants appreciated that it encouraged them to move with their dog in order to play, e.g., \textit{``I like that it keeps us both active [...], I wasn't just sitting down [...] I really liked [that] we were both active during the game''} (TT-P7) (see Fig. \ref{fig:moving_together}). 

\subsection{Affordances of Augmentation}

\subsubsection{Entertaining to watch (39 participants).} 
\label{sec:entertain}
\begin{figure}[] \centering \includegraphics[width=\linewidth]{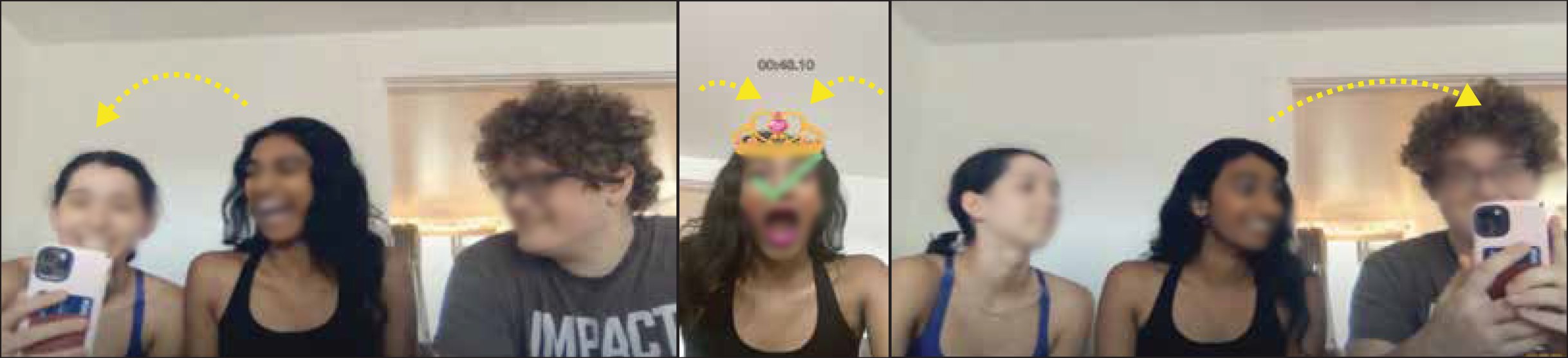} \caption{\label{fig:nonverbal_immediacy} FI participants were engaged and entertained even when others held the phone.} \end{figure}

Augmentations can be entertaining. When they are connected to other players' bodies, they had funny or surprising qualities. This made the experience enjoyable to watch. For example, the continuously changing face augmentation in Face It surprised and delighted participants, e.g., \textit{``I like that it's very random [...] when the filters change [...] it's kind of like a fun element to it. I really enjoyed that''} (FI-P1). The design of the augmentations themselves also contributed to that excitement, e.g., \textit{``At first, I was like, 'Oh, these filters are kind of [laughing] kind of out there.' But I think that also makes it more exciting [...] you know, it's just, it's wacky, it's funny''} (FI-P20). Participants mentioned this made the experience \textit{``fun''} and \textit{``funny''}, e.g., one participant observed that \textit{``adding that [AR] filter definitely makes it more uh, interesting and fun to play''} (FI-P20), while another noted that when they were using Face It, it was \textit{``pretty entertaining and fun to watch other people make faces''} (FI-P16) (see Fig. \ref{fig:nonverbal_immediacy}). Treasure Treat participants enjoyed watching their dogs collect the AR coins, e.g., \textit{``I was excited while [laughing] I was using like, this is super cool. Uh, I was smiling the whole time watching her win through the rounds''} (TT-P7), and \textit{``I would record a video of her, like her tail wagging and catching a coin [...] I think that's adorable''} (TT-P5).

\subsubsection{Supporting experimentation together (31 participants).} \label{sec:experiment}
\begin{figure}[] \centering \includegraphics[width=0.7\linewidth]{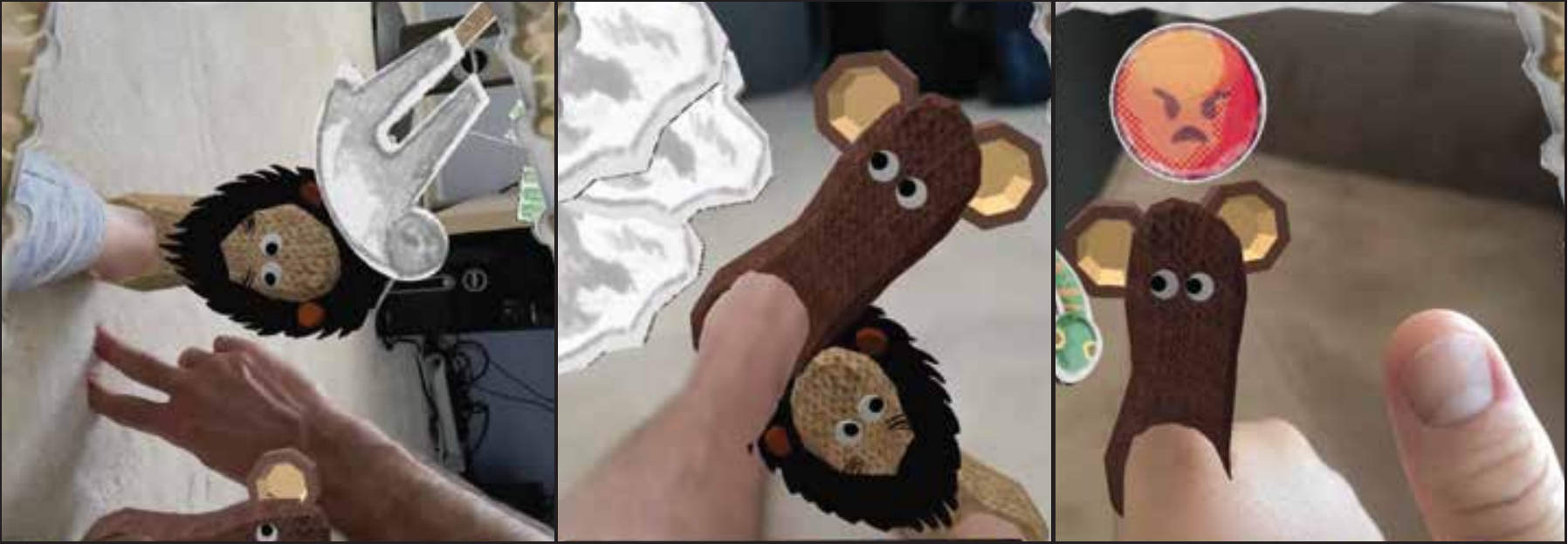} \caption{\label{fig:experimented} Parents and children used FF in unexpected ways, e.g., tilting their phones and playing with their hands in the frame.} \end{figure}

Augmentations can be surprising, and they can contribute to the enjoyment when they're designed to be discovered as the experience unfolds (rather than all at once). In addition, we found that when participants experiment together to discover augmentations it adds to the fun. For example, augmenting feet into sock puppet animal characters was part of the fun discovery of Feature Films (e.g., \textit{``her favorite thing was just being able to turn her feet into animals''} (FF-P20)). We also observed participants playfully experimenting with the AR tracking, e.g., one pair of participants used their thumbs instead of their feet. Participants also explored adding finger characters to the mix (e.g., \textit{``I added my little finger guy in there and tickled him a little''} (FF-P3)) (see Fig. \ref{fig:experimented}).

However, participants indicated that \textit{``the best part''} (FF-P1) was \textit{``being able to engage with it together''} (FF-P1). Many Feature Films participants commented they wanted to be continuously surprised and able to experiment with new content. They suggested adding a variety of changing stories, backgrounds, settings, and characters. They found it desirable and thought it could increase the replay-ability of the app, e.g., one participant stated, \textit {``if there was a different story, like every week or every day, I would go back and use [it], as long as there's a different story''} (FF-P1). Participants also experimented while using Treasure Treat; some tried different approaches to having their dog's body collect the AR coins. For example, instead of encouraging the dog to move to collect AR coins, they moved around themselves. That made the dog's body collide with the AR coins to collect them, e.g., \textit{``I could basically hack the system and like cover Rocky with a coin, and then we would get the points''} (TT-P6). In Freezing Frenzy, some participants experimented by trying to augment additional body parts rather than simply using their feet, e.g., \textit{``not being able to, like, put ice on her face. It always went to her head, and I kept trying''} (FR-P11).

\subsection{Co-Located Play}
\subsubsection{Enhancing friendly competition (41 participants).} \label{sec:competition}
Participants who played our competitive apps (Face It, Freezing Frenzy, and Milky Way) mentioned that they experienced the competition as friendly and that engaging in these digital experiences in person rather than remotely enhanced them. For example, interacting with the apps \textit{``actually like live''} made them \textit{``more interactive''} (FI-P2) and engaging. Face It fostered a fun sense of thrill through competition, e.g., \textit{``the speeding up of the music as time went on, like it got more and more urgent. I thought that was really fun''} (FI-P3). Being able \textit{``to see the person [...], look at them, do the competition right [t]here in the same space''} (FI-P10) enhanced the competition. In Face It, the core game mechanic of \textit{``passing the phone with each other,''} (FI-P14) was identified as \textit{``more of a physical thing that you do''} (FI-P14) and \textit{``remotely [it] wouldn't be the same since it's such a fast-paced game''} (FI-P9). \textit{``[P]laying a game with someone right next to you''} (FI-P7) was something participants appreciated and considered \textit{``unique''} (FI-P3). When playing Freezing Frenzy and Milky Way comparing scores at the end of each game round provided an opportunity for players to connect. Otherwise, they would not have known what the other person was up to, e.g., \textit{``we would go back over and be like, "what was your score?"''} (FR-P12). 

\begin{figure}[] \centering \includegraphics[width=1\linewidth]{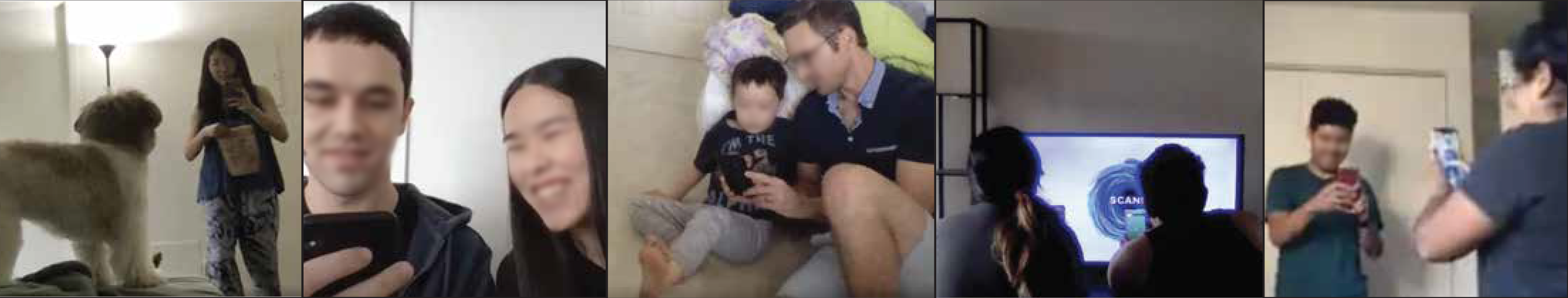} \caption{\label{fig:enaged_actively} All five apps engaged participants actively with others.} \end{figure}

\subsubsection{Making room for more (55 participants).} \label{sec:room} 
Mobile AR apps can be designed for multiple players, and our participants appreciated the ability to play with \textit{multiple co-located people} and to allow more people to join them. Face It participants also mentioned the game is \textit{``more of a party game''} (FI-P15), which would be \textit{``more fun with multiple people''} (FI-P15). One participant discussed the way sharing one phone with the mechanic of passing it around was what made it more engaging \textit{``because you can play with multiple people''} (FI-P1). Many Freezing Frenzy participants also imagined how the use of individual phones could support a larger group of players, e.g., \textit{``it'd be really interesting to just have a bigger group of people''} (FR-P9). Participants who played Milky Way thought the second large screen would allow more players to be added, e.g., \textit{``I feel like the TV is for like, multiple people in the same room''} (MW-P8). MW-P7 discussed the advantage of using the app with \textit{``bigger screens and movie theaters''}, envisioning the potential for \textit{``hundreds of people''} to play it simultaneously. Interestingly, even small screens were able to achieve this party effect with the Face It app.

\subsubsection{Making screen time interactive (43 participants).} \label{sec:interactive}
All of the apps engaged participants actively with others on some level (see Fig. \ref{fig:enaged_actively}). For example, participants felt that Face It could work as a kind of ice-breaker (e.g., \textit{``it might be fun, like in the car or something. You know, [on a] long road trip. You're just trying to bond with your fellow travelers''} (FI-P20)). For Treasure Treat participants, the app motivated them to engage with their dogs. One participant said, \textit{``I have a reason to try and get my dog to do something. And it's an external goal that we're trying to achieve together''} (TT-P2); this participant mentioned that they thought their dog \textit{``felt like there was more play happening than, than just me tossing a toy away and not caring what the outcome was. It felt like we were playing''} (TT-P2). Participants discussed how the apps made \textit{``screen time”} social and interactive. For example, one participant shared this comment about Feature Films: \textit{``it's not like watching a TV show where it's totally passive; It's something you do together. It's like a good five-minute activity for you and your kid''} (FF-P3). Another participant discussed the app's benefits in terms of supporting them to be playful with their child: \textit{``it's like a little way to get some of the sillies out [...] knowing that it's like a share-together activity [and] to not feel guilty over that screen time''} (FF-P20). Participants also discussed the way having the apps ready on their phones could make it easier for them to initiate the interactive experience with others. For example, a Feature Films participant mentioned that using the app on their phone made it more readily available for them to use with their child than reading a storybook, observing \textit{``there's an ease to it where you don't need a lot. You don't have to go pick a book off your bookshelf''} (FF-P5).

\section{Discussion}
Earlier, we argued that mobile AR is well suited to supporting co-located interactions because it is grounded, embodied, playful, social, and memorable. We then identified a set of design attributes that guided our development of the apps: (1) device arrangement, (2) enablers, (3) augmentation, (4) interaction type, and (5) movement. 

\subsection{Key Takeaways}
Based on learning from our synthesis of existing articles in the literature review (\hyperref[sec:lit]{Section 2.3}) and our user study (\hyperref[sec:results]{Section 5}), we present a set of guidelines that designers can use in making choices related to these design attributes, which are focused on the four main categories (see Table \ref{fig:IRL-Tables-Takeaways}). We developed these guidelines after completing the user study, and they include our design-focused reflections on the process. The underlying questions guiding their development were as follows: how does this experience truly leverage co-location, and would this be nearly impossible to replicate at a distance?

\textbf{1. Device Arrangement.} How are mobile phones distributed among participants? For example, does each person hold their phone? Do they pass one phone around? Or something else? In making this decision, designers should consider at least two approaches to shaping the experience:
\begin{itemize}
    \item \textit{Encouraging touch through proxemics.} Create a digital experience that gets people close to one another so they can touch, for example, by playfully bumping and pushing each other. Physical human contact makes co-location valuable because it is hard to replicate in remote interactions (related sections: \hyperref[sec:lit]{2.3--Emobdied; Social}, \hyperref[sec:proxemics]{5.1.1}).
    \item \textit{Nudging person-to-person communication via coordination.}
    Create playful scenarios that require people to share information. For example, introduce collaborative tasks that nudge people to talk to one another, especially using their voices, in order to complete them (related sections: \hyperref[sec:lit]{2.3--Emobdied; Playful; Social}, \hyperref[sec:coordination]{5.1.2}).
 
\end{itemize}

\textbf{2. The Roles of Enablers.} How are visual triggers used to foster social interaction? AR systems often relay on visual markers to function. Designers should explicitly consider the potential for markers to play a social role. 
\begin{itemize}
  \item \textit{Use bodies as enablers.} Steer people's attention towards one another by using their faces, feet, arms, and full bodies as enablers. This could help to make people, not technology, the focus of social interaction. In addition, consider using body tracking to nudge people to move together (related sections: \hyperref[sec:lit]{2.3--Grounded; Emobdied; Social}, \hyperref[sec:bodies]{5.2.1}).

 \item \textit{Lean on physically meaningful enablers.} Rather than using meaningless codes as enablers, embrace the value of physical enablers. For example, using pets as enablers leverages our physical world because, despite advances in telepresence, pets typically respond best to physical touch, smells, and other analog signals. Similarly, using enablers from people's built environment (e.g., objects in a living room) helps ground the social interaction in their shared space and can also support movement (related sections: \hyperref[sec:lit]{2.3--Grounded; Emobdied}, \hyperref[sec:move]{5.2.2}).
    
\end{itemize}

\textbf{3. Affordances of Augmentation.} AR experiences involve visually altering the user's view of the physical world. Designers can use these alterations to support co-location in at least two ways. 
\begin{itemize}
 \item \textit{Entertaining both the player and others.} Utilize the visual effects of the augmentations as comedic devices for disarming people with laughter, for example, by turning the user's face into a potato. In addition, changing the visual effects throughout a game session can help keep the experiences novel and engaging (related sections: \hyperref[sec:lit]{2.3--Playful; Social}, \hyperref[sec:entertain]{5.3.1}).
 
\item \textit{Fostering opportunities to experiment together.} Use augmentations that rely on people moving around and pointing their cameras at different parts of the physical space to allow them to explore and experiment with others in their environment (related sections: \hyperref[sec:lit]{2.3--Grounded; Playful; Social}, \hyperref[sec:experiment]{5.3.2}).

\end{itemize}

\textbf{4. Co-located Play}. Although game design is a field on its own, we identified three ways designers might want to optimize technologies for co-location.
\begin{itemize}
 \item \textit{Leverage play “in real-time”.} Whether the playful activity is competitive or cooperative, finding the right balance between fun and challenge also involves attention to physical dexterity. Further, designers should consider making the level of challenge relate to people's physical skills, as physical skills are more aligned with co-location than non-physical cognitive skills (related sections: \hyperref[sec:lit]{2.3--Embodied; Playful; Social}, \hyperref[sec:competition]{5.4.1}).

\item \textit{Enable any number of players.} Allow people to try out the experience on their own first, but show them how adding more players would make the experience more fun. Design interactions so they can be easily set up to make room for the active involvement of multiple players (related sections: \hyperref[sec:lit]{2.3--Grounded; Playful; Social}, \hyperref[sec:room]{5.4.2}). 
 \item \textit{Build on familiar play.} Draw inspiration from existing game experiences people might already be familiar with in the analog world, e.g., board-games and toys. This could help people get started quickly while still exposing them to novel game techniques (related sections: \hyperref[sec:lit]{2.3--Grounded; Embodied; Playful}, \hyperref[sec:interactive]{5.4.3}). 
\end{itemize}

\subsection{Embodied Social AR}
When we designed the IRL apps, we prioritized taking into consideration the context where we envisioned people would use the app. The \textit{interaction type} design attribute guided us to explore a variety of use contexts carrying a range of social dynamics-- from collaborative storytelling or working together with one's pet-- to competitive play experiences among friends.As a result, it also affected the embodied social experience. For example, when designing Feeture Films we thought about where and how parent-child interactions occur. This user context guided us to create a collaborative experience while sticking one's feet out because that's what parents and children often do (while laying on a bed or a couch together).

However, the \textit{enabler} design attribute became particularly generative and important during the brainstorming stage of the design process. Enablers are not just physical entities that trigger the AR experience--they also play an integral role as the central augmentation from which the entire experience unfolds. Brainstorming around the enablers allowed us to ground the experiences in the physical reality of players. The process of designing and studying the apps with participants led us to notice that there was value in focusing on enablers that augmented bodies (e.g., faces, feet, dogs, and the full human body). 

The core idea of \textit{embodied social AR} suggests that social AR experiences can benefit from focusing on enablers that create embodied interaction, in particular, grounding the enablers in other social beings to trigger and drive interactive experiences and, in the process, explore the potential of the whole body and its affordances for play and movement. Therefore, embodied social AR can encourage experiences that include movement and bodily play, and focus players' attention on one another. This enables people to (1) trigger the experience by detecting a body, (2) the body detected is then being augmented, and (3) the results of the previous points enhance the co-located social experience of the players. In other words, tracking bodies is the input that initiates the interaction (i.e., button press). Augmenting those bodies is the central output (i.e., the result), while the outcome impacts the players' social experience. We believe embodied social AR will only grow in popularity, and like any other type of design material, it should be used for game design.

\begin{table}[t]
\footnotesize
\renewcommand{\arraystretch}{1.3}
\caption{Design recommendations: Key takeaways.}
\label{fig:IRL-Tables-Takeaways}
\centering
\footnotesize

\begin{tabular}{p{0.27\textwidth}p{0.29\textwidth}p{0.35\textwidth}}
\toprule
&\multicolumn{2}{c}{\bfseries Design Recommendations}\\\cmidrule{2-3}
\textbf{Device Arrangement} & Encourage touch through proxemics & Needing to coordinate could nudge communication \\
\textbf{The Roles of Enablers} & Center around players' bodies & Lean on physically-meaningful enablers \\
\textbf{Affordances of Augmentation} & Entertain both the player and others & Foster opportunities to experiment together \\
\textbf{Co-located Play} & Leverage play ``in real time'' & Easily to set-up, let multiple players join in \\
\bottomrule
\end{tabular}
\end{table}

\subsection {Future Work}
The five IRL apps represent our initial exploration, and as design instances, they only reflect aspects of the potential design space for playful co-located mobile AR. There are many other ways to explore it:

\begin{itemize}
    \item \textbf{Different combinations of design attributes.} Experiments combining the same design attributes we explored in different ways would yield other designs that could be fun to play.
    \item \textbf{Changing details of the design attribute.} Researchers could use the same design attributes we explored but change their specifics. For example, using furniture as an enabler in place of bodies for a shooter game and studying its effect on the co-located experience.
    \item \textbf{Experimenting with other design attributes.} Experiment with design attributes other than those we explored. For example, familiarity with previous related experiences; augmentation based on the front vs. the rear camera view; dimensions of the co-located space; indoor vs. outdoor; and more. 
    \item \textbf{Exploring embodied social AR.} Study the effects of enablers that focus specifically on augmenting bodies to unpack the value of embodied social AR.
    \item \textbf{Focusing on specific relationships.} Explore specific types of relationships, such as inter-generational, peers, parents, grandparents, co-workers, etc. 
    \item \textbf{Studying interaction with pets.} It would be interesting to study co-located interaction with pets (other than dogs) and interactions between multiple people and one pet.
     \item \textbf{Considering the effects of networked backend or the lack of it.} Compare and study the impact of having an available networked backend vs. not having one would be valuable (to determine whether players can have shared access to the same game elements on their devices or not). 
     \item \textbf{Determining the potential for re-playability.} This includes studying how we could create playful co-located mobile AR interactions that are just as enjoyable when played over and over again (i.e., finding ways to infuse them with the qualities that makes board games enjoyable to play and re-play).
      \item \textbf{Exploring the effects of memorability.} Mobile AR technology could be a great design material for creating memorable shared experiences. Our study represents a thin slice in time, so we could not test the impacts of the five apps in terms of the memorability of the co-located experiences they facilitate. A more longitudinal approach is required to assess the memorability of playful co-located AR experiences. 
     \item \textbf{Developing a richer taxonomy for mobile AR.} In our exploration, we faced the need to discuss the nuances of what triggers AR experiences and their effects. Therefore, we created our taxonomy using the term \textit{enabler} to describe the physical entities that trigger and become the focus of AR experiences. In analyzing our results, we became aware, once again, of the lack of appropriate terminology for discussing different types of augmentations (e.g., face morphing, customizing, filters, object additions, overlays, etc.). Prior work has begun to discuss some of these nuances \cite{hugues2011new, Normand2012TypologyAR}. However, more work is needed to develop richer terminology for interaction design with mobile AR (while also taking into account co-location) to allow for more fruitful discussion of its affordances. 
\end{itemize}

\section{Limitations}
If not for the restrictions imposed by the COVID-19 pandemic, we would have evaluated the playful co-located mobile AR apps with participants in person. This study would have been easier to run onsite rather than remotely, and it may have resulted in additional findings about the players' social experiences. However, conducting the study remotely did not affect the quality of our results. Our participants still engaged with our apps while co-located with others, and we extracted insightful design takeaways. 

In terms of the IRL apps generated in the study, we explored only a subset of design possibilities and focused solely on mobile AR. Therefore, it is likely that our findings do not represent a comprehensive analysis of the potential for co-located interaction. For example, we did not explore how devices other than mobile phones (e.g., smart speakers) could enhance or encourage co-located playful interaction, nor did we consider gender and age differences in our analysis. Moreover, we experienced some usability issues at times. However, we did not detail them as we felt that they did not add any relevant insights to our discussion.

\section{Conclusion}
In this paper, we presented project IRL: a suite of five apps designed for playful co-located interaction using mobile AR. We took a research through design approach \cite{RtDZimmerman, Gaver2012RTD} to learning and exploring how mobile AR can enhance players' co-located social experiences. We designed and deployed the apps with 101 participants. From our results, we identified design insights that we organized under four themes: (i) device arrangement, (ii) enablers, (iii) augmentations, and (iv) co-located play. For each theme, we developed design recommendations as the key takeaways (Table \ref{fig:IRL-Tables-Takeaways}). We learned that opportunities exist to enhance the co-located experience with playful mobile AR by engaging players in interactions that encourage them to move together or by creating a social focal point. The device arrangement (e.g., shared phone vs. use of multiple phones) can further impact the social experience by shaping proxemics or encouraging communication for coordination between players. Augmentations could support playful co-located interaction by engaging players through funny and curious situations. Mobile AR can serve as an effective additional design material that creators of co-located play experiences can use to facilitate friendly competitions, to engage multiple players, and to make `screen time' more social, entertaining, and interactive. 

When we connected AR to bodies as enablers, our apps supported co-located play by encouraging participants to move together and focus on each other more. We hope to inspire future work that focuses on designing playful co-located mobile AR. We suggest following the \textit{embodied social AR} approach to generate engaging designs for co-located play experiences and to explore different enablers (however, such research should by no means be limited to bodies only). We also believe transforming screen time into a shared and active experience is one way to recover social engagement in home environments. We hope designers and researchers continue to challenge existing technologies by designing interactions that help people re-connect. 

\bibliographystyle{ACM-Reference-Format}
\bibliography{IRL-arxiv}

\appendix

\section*{Appendix}
\section{App Access Codes} \label{sec:snapcodes}
The co-located AR apps developed for IRL can be accessed using Snapchat \cite{snap}, you can follow the instruction on Snapchat's \href{https://lensstudio.snapchat.com/guides/sharing/unlocking-lenses/}{official documentation}.

\begin{enumerate}
    \item Open Snapchat 
    \item Point the camera at the images provided on Fig.\ref{fig:snapcodes} 
    \item press and hold your finger on top of the image
    \item Select unlock for 24/48 hours
\end{enumerate}

\begin{figure}[h]
  \centering
  \includegraphics[width=0.25\linewidth]{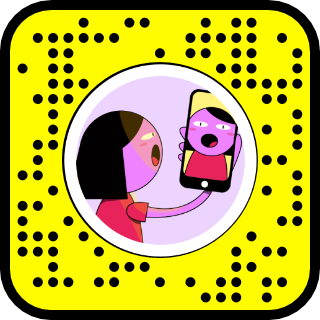}
    \caption{\label{fig:snapcodeFF} \textit{``Snapcodes"} for Face-it (FI). }
\end{figure}

\begin{figure}[h]
  \centering
  \includegraphics[width=0.25\linewidth]{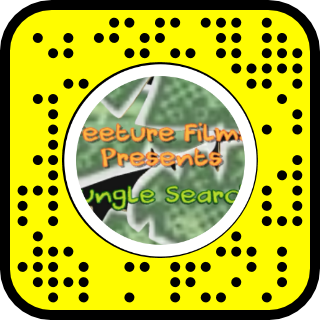}
    \caption{\label{fig:snapcodeFF} \textit{``Snapcodes"} for Feeture Film (FF).}
\end{figure}

\begin{figure}[h]
  \centering
  \includegraphics[width=0.25\linewidth]{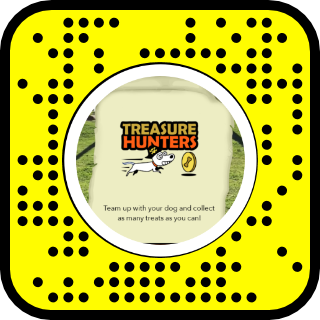}
    \caption{\label{fig:snapcodeFF} \textit{``Snapcodes"} for Treasure Treat (TT).}
\end{figure}

\begin{figure}[h]
  \centering
  \includegraphics[width=0.25\linewidth]{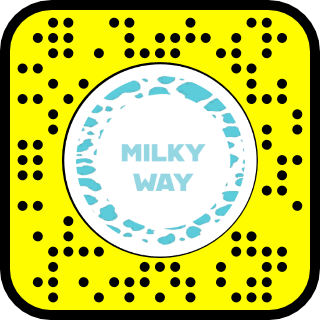}
    \caption{\label{fig:snapcodeFF} \textit{``Snapcodes"} for Milky Moo (MW).}
\end{figure}

\begin{figure}[h]
  \centering
  \includegraphics[width=0.25\linewidth]{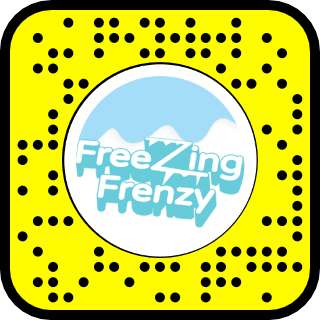}
    \caption{\label{fig:snapcodeFF} \textit{``Snapcodes"} for Freezing Frenzy (FR).}
\end{figure}

\newpage

\section{Semi-structured Interviews}

The following is a list of the potential questions we asked in our semi-structured interviews

\begin{itemize}
	\item What was your first impression? OR What were your thoughts about what you just experienced?
	\item How would you describe this experience to your friend who has never played it before? What would you tell them about it?
    \item Now that you have had a chance to experience it, is there any information that would have been useful to you before starting?
	\item Was anything confusing? (Please take me through what you found to be confusing) 
	\item Was there anything you found particularly frustrating?
	\item What did you find surprising about experience? 
	\item What do you think is special about this experience? OR what are the things that make it different from other things you’ve used before?
	\item What were the main challenges for you? OR What were the most challenging things for you with this experience?
	\item What did you like most about the experience?
        \begin{itemize}
	        \item If it was fun, what was fun about it?
	        \item If it wasn’t enjoyable: what didn't work for you? What you wish would be different?
        \end{itemize}
	\item What did you enjoy the least when using it?	
	\item What was the most exciting moment?
	\item In what way did you interact with other people (or with your dog)?	
	\item Do you prefer to use it alone or with other people or with your dog)?
	\item In person vs online playing
	      \begin{itemize}
	        \item Comparing it to playing in-person vs. playing something similar on-line or other `similar' online games.
	        \item Example question: how do you think using the lens like you did, in person is different than using it online with someone?
        \end{itemize}
	\item Physical interaction
	\begin{itemize}
	        \item Questions about the device configuration (sharing one phone/using different phones/ including a video on TV)
	        \item Questions related to interaction through required body movements (sharing one phone/using different phones/ including a video on TV)
        \end{itemize}
    \item Social interaction
    	\begin{itemize}
	        \item What do you think about using it with other people vs. playing it alone?
	        \item What did you think of the `type of interaction' (e.g. the competition, using something to tell a story, playing a levels game)
        \end{itemize}
	\item Recording
		\begin{itemize}
	        \item (if it recorded) What do you think of the fact that it recorded you using it, and then you were able to save it? 
	        \item Would you look at it again? 
	        \item Would you want to send it to someone else? 
	        \item Who would you share it with? Why?
	        \item Would you want to record the experience?
        \end{itemize}
	\item Sharing
		\begin{itemize}
	        \item What if you could share this experience with someone-- who would you share it with?
	        \item Would you like to share this experience with other people? 
	        \item How would you share it? 
	        \item Would you send videos to people? Post to on social media? 
	        \item Would you talk about with other people?
        \end{itemize}
	\item Duration and pace
		\begin{itemize}
	        \item What about the timing, the length of the experience?  
	        \item How did it feel-- did it feel too long, too short, or just about right? 
	        \item What did you think of the pace of it?
        \end{itemize}
	\item Demo video	
	\begin{itemize}
	        \item What did you think when you watched the demo video?
	        \item What did you think about the video? 
	        \item Did watching it affect how you interacted?
        \end{itemize}
	\item How did the experience change when you used it for the second time?
	\item What elements do you think could be improved? 
	\item Is there anything that you did not like about the experience? What for example?
	\item Is there anything that you thought I would ask but I didn’t? Anything you wanted to talk about?	
	\item Overall, how would you describe this experience’s appeal?
	\item What parts of the experience attracted you? OR what parts of the experience you thought worked particularly well?
	\item What was missing from the experience?	
	\item What if you could change just one thing, what would it be?	
	\item How was it to learn how to use it?
	\item Would you purchase it?		
\end{itemize}

\end{document}